\documentclass[preprint,...]{revtex4}

\usepackage{amsmath, amssymb, appendix, graphicx}
\usepackage{pdfpages}

\newcommand{\be}{\begin{equation}}
\newcommand{\ee}{\end{equation}}
\newcommand{\bea}{\begin{eqnarray}}
\newcommand{\eea}{\end{eqnarray}}

\newcommand{\bef}{\begin{figure}}
\newcommand{\enf}{\end{figure}}
\newcommand{\p}{\mathbf{p}}
\renewcommand{\r}{\mathbf{r}}

\newcommand{\x}{\mathbf{x}}
\newcommand{\y}{\mathbf{y}}

\newcommand{\A}{\mathbf{A}}
\newcommand{\w}{\omega}

\begin{document}

\title{Time-resolving electron-core dynamics during strong field ionization in circularly polarized fields}

\author{Lisa Torlina}
\affiliation{Max Born Institute, Max Born Strasse 2a, 12489, Berlin, Germany}

\author{Jivesh Kaushal}
\affiliation{Max Born Institute, Max Born Strasse 2a, 12489, Berlin, Germany}

\author{Olga Smirnova}
\affiliation{Max Born Institute, Max Born Strasse 2a, 12489, Berlin, Germany}
%\normalsize{$^{\ast}$These authors have contributed equally.}
\date{\today}

\begin{abstract}
Electron-core interactions play a key role in strong-field ionization and the formation of
 photoelectron spectra. We analyse the temporal dynamics of strong field ionization associated with these interactions using the time-dependent analytical R-matrix (ARM) method, developed in our previous work [J. Kaushal and O. Smirnova, Phys. Rev. A 88, 013421 (2013)]. The approach is fully quantum but includes the concept of trajectories.
However, the trajectories are not classical in the sense that they have both real
 and imaginary components all the way to the detector.
We show that the imaginary parts of these trajectories, which are usually ignored,
 have a clear physical meaning and are crucial for the correct description of electron-core interactions after ionization. In particular, they give rise to electron deceleration, as well as dynamics associated with electron recapture and release.
 Our approach is analytical and time-dependent, and allows one to gain access to the electron energy distribution and ionization yield as a function of time.
Thus we can also rigorously answer the question: when is ionization completed?

\end{abstract}
%\title{Non-adiabatic Coulomb effects in strong field ionisation by circularly polarised fields}

\maketitle
\section{Introduction}

We address the effects of electron-core interaction in strong field ionization by circular fields. 
These effects play a crucial role in the dynamics of electron release and hole formation by strong field ionization. 
They determine the structure of photoelectron spectra, ionization delays, and the effects of electron re-capture into bound states.
Understanding these effects is one of the key components of attosecond imaging. 

In addition to the direct ab-initio solution of the time-dependent Schroedinger equation (TDSE) (e.g. \cite{muller,Madsen2010}), standard approaches to calculating photoelectron spectra in strong field ionization rely on the two-step model. The first step involves ionization in a static field. This step is treated within the quantum approach and is often viewed as the electron tunneling through the barrier created by the  laser field and the core potential. The second step describes electron dynamics after ionization and is treated classically by propagating classical trajectories from the ``exit'' of the tunneling barrier to the detector. A more advanced approach (CCSFA) involves classical Monte-Carlo-type simulations, with the additional capability of treating non-adiabatic effects during ionization,  keeping track of phases accumulated along each trajectory and treating dynamics inside and outside the barrier on an equal footing by using complex times \cite{poprchapt,poprprl2010,holography}. Using this method, good agreement between theory and experiment has been demonstrated for elliptically polarized fields \cite{popr2008}.
 The appeal of these models is in the accessibility of the physical picture. However, the difficulty lies in separating the process into quantum and classical steps.
For the case of a long-range electron-core interaction, the two-step model does not provide a unique recipe for merging the quantum and classical treatments.

We overcome these difficulties in our approach, which provides a consistent quantum treatment, yet includes the concept of trajectories. However, in our case the trajectories are not classical in the sense that they have both real and imaginary components all the way to the detector. The imaginary components of trajectories outside of the barrier are usually excluded by re-defining the initial conditions prior to ionization \cite{jmopopr,bauer2012}. Here, we show that the imaginary parts are in fact responsible for a number of effects, including electron recapture and re-release during ionization, and the redistribution of electrons over continuum states after ionization, which corresponds to their deceleration due to the interaction with the core.

We investigate the effect of the Coulomb correction on ionization amplitudes in circularly polarized fields using results obtained within the formalism of the time-dependent analytical R-matrix \cite{ARM1,ARM2,circARM1}.
The time-dependent analytical R-matrix (ARM) method provides an opportunity to develop a consistent theory of
 strong-field ionization for arbitrary core potentials in the time domain.
In the spirit of R-matrix theory, the configuration space is split into two regions:
 an inner region enclosing the atom or molecule, and an outer region, remote from the singularity of the core.
The remotness of the outer region allows one to use propagators based on the eikonal-Volkov (EVA) states.
 The latter are known to perform well for soft-core potentials \cite{Smirnova08} and to allow one to introduce 
 and analyse the sub-cycle ionization yield \cite{Smirnova06}.
The matching of inner region solutions to outer-region EVA states occurs at the R-matrix boundary $a$, which
 is placed in the asymptotic region $a>>1/\kappa$, where $1/\kappa$ is the size of the ground state.
 Technically, matching is achieved using the Bloch operator. 
 
In \cite{circARM1}, we derived an expression for ionization amplitudes in circularly polarized fields using the ARM approach, and analysed the effect of the long-range electron-core interaction on ionization times, the initial conditions for electron continuum dynamics and the ratio of ionization rates from $p^-$ and $p^+$ orbitals. Here, we focus on non-adiabatic Coulomb effects in photoelectron spectra and investigate subcycle variations in the total ionization rate. This analysis enables us to gain insight into the temporal dynamics of ionization, allowing us to determine when ionization is completed and to explore how this depends on the laser parameters.
 
%  In \cite{circARM1}, we analysed
%  the effect of long-range electron-core interaction on the ionization rates.  
%Here, we focus on non-adiabatic effects in photoelectron spectra, which appear for Keldysh parameter  $\gamma>~1$.
%We also analyse sub-cycle ionization yield and sub-cycle photoelectron spectra. This analysis allows us to gain
% insight into temporal dynamics of ionization to see when the ionization is completed  and how its completion depends on the laser parameters.

This paper is organized as follows. In section II we present and discuss the expression for the ionization amplitude derived in our previous work \cite{circARM1}. In section III, we introduce the concept of electron trajectories and analyse the Coulomb correction term in detail. In section IV, we investigate the resulting effects on electron spectra and time-resolved ionization rates. Section V concludes the paper.

\section{Ionization amplitudes for circular fields}\label{sec:amplitudes}

The relevant quantity for studying ionization rates and electron spectra is the ionization amplitude, $a_\p(T) = \langle \p | \Psi(T) \rangle$, where $|\Psi(T)\rangle$ is the wavefunction of the liberated electron at some time $T$, after the laser field has been switched off. An analytic expression for this was derived in \cite{circARM1} using the ARM approach.

In particular, if our field is given by $\mathbf{E} = F(-\sin(\w t) \ \mathbf{\hat{x}} + \cos(\w t) \ \mathbf{\hat{y}})$, with corresponding vector potential
\begin{equation}
	\mathbf{A} = -A_0 (\cos(\w t) \ \mathbf{\hat{x}} + \sin(\w t) \ \mathbf{\hat{y}}),
\end{equation}
the expression for the ionization amplitude takes the following form:
\begin{equation}
	a_{\mathbf{p}}(T) = \langle \p | \Psi(T) \rangle = R_{\kappa l m}(\p) \ e^{-i W_C(T,\p)} \ e^{-i S_{SFA} (T,\p)}.
\end{equation}
Let us briefly review each part of this expression in turn.

First, $e^{-i S_{SFA}}$ is the standard exponential factor that arises in the strong field approximation. Expressed in terms of dimensionless time $\phi_T = \w T$,
\begin{equation}
	S_{SFA}(\phi_T,\p) = \frac{1}{2\w} \int_{\phi_s}^{\phi_T}[\p+\A(\phi)]^2 d\phi - \frac{I_p}{\w} \phi_s.
\end{equation}
Here, $I_p$ is the ionization potential, and $\phi_s = \phi_i + i \phi_\tau$ denotes the complex solution to the saddlepoint equation $[\p+\A(\phi_s)]^2 = -2I_p$. Explicitly, we have
\begin{align}
	&\phi_i = \omega t_i = \tan^{-1} \left(\frac{p_y}{p_x}\right), \\
	&\phi_\tau = \w \tau_T = \cosh^{-1} \left[\frac{A_0}{2 p} \left(\gamma_{\mathrm{eff}}^2 + \frac{p^2}{A_0^2} +1 \right) \right],
\end{align}
where $p=\sqrt{p_x^2 + p_y^2}$, $\gamma_\mathrm{eff} = \omega/F\sqrt{\kappa^2 + p_z^2}$ is the effective Keldysh parameter, and $\kappa = \sqrt{2 I_p}$. We interpret $t_i$ as being the ionization time, that is, the time at which the electron emerges in the continuum. Notice that here $\phi_i=\omega t_i$ corresponds to the angle at which we detect our electron. This result neglects corrections to the ionization time (for a given observation angle) due to the core potential: taking Coulomb interactions into account, the ionization time is modified to $\phi_i = \phi_i^{(0)} - |\Delta \phi_i| = \tan^{-1}\left(\frac{p_y}{p_x}\right) - |\Delta \phi_i|$ (see \cite{circARM1} for details). To first order, however, these corrections only affect the prefactor $R_{\kappa l m}(\p)$ and do not appear in the exponential terms. Since we will focus on the exponential terms in this work, we can neglect these corrections here. Note, also, that we select only a single solution for $\phi_s$, which is associated with a single ionization burst. 

For a given ionization angle, $|e^{-i S_{SFA}(p)}|^2$ is a Gaussian-like distribution in momentum, centred at optimal momentum $k_0$, where
\begin{equation}\label{eq:k0}
	k_0 = A_0 \sqrt{1+\gamma^2} \sqrt{\frac{1-\zeta_0}{1+\zeta_0}},
\end{equation}
and the parameter $0 \leq \zeta_0 < 1$ satisfies $\sqrt{\frac{\zeta_0^2 + \gamma^2}{1+\gamma^2}} = \tanh\left(\frac{1}{1-\zeta_0} \sqrt{\frac{\zeta_0^2 + \gamma^2}{1+\gamma^2}} \right)$ \cite{Barth2011}. In the adiabatic $\gamma \ll 1$ limit, $k_0 \rightarrow A_0$. As we shall see, this corresponds to the electron appearing in the continuum with zero initial velocity. In general, however, $k_0 > A_0$ and the electron is born with an initial velocity in the direction in which the barrier is rotating. 

The next term, $e^{-i W_C}$, represents the Coulomb correction coming from the interaction between the departing electron and the core. It is given by
\begin{equation}
	W_C(\phi_T,\p) = \frac{1}{\omega} \int_{\phi_i + i\phi_\kappa}^{\phi_T} d\phi \ U(\r_s(\phi)), \label{eq:WC}
\end{equation}
where $U(\r_s)$ is the core potential of the atom or molecule evaluated along the trajectory of the departing electron,
\begin{equation}
	\r_s(\phi) = \frac{1}{\w} \int_{\phi_i + i \phi_\tau}^{\phi} d\phi' \ (\p + \A(\phi')),
\end{equation}
and $\phi_\kappa = \phi_\tau - \omega/\kappa^2$. Understanding this term and its effect on ionization rates and electron spectra will form the core of this paper. Notice, for now, that $W_C$ is both momentum and time dependent. This will allow us to study the effect of the core potential on electron spectra and to probe the process of ionization in a time-resolved way.

The final term, $R_{\kappa l m}(\p)$ encodes the angular structure of the initial state. Since the aim of this paper is to investigate the role of the Coulomb correction term $e^{-iW_C}$, we shall focus on the simplest case of the spherically symmetric state with no angular momentum $l=m=0$. In this case, $R_{\kappa l m}(\p)$ reduces to $R_{\kappa 0 0}(\p) \propto \sqrt{\kappa}/\sqrt{S_{SFA}''(t_s^{(1)}(\p))}$, where $t_s^{(1)}$ is the Coulomb-corrected ionization time. The overall effect of this prefactor is to slightly shift the electron momentum distribution to lower momenta. We shall neglect this effect here. The case $l=1$ was considered in \cite{circARM1}, where we investigated the impact of non-adiabatic Coulomb effects on the sensitivity of ionization to the sense of electron rotation in the initial state, originally predicted for short-range potentials in \cite{Barth2011}.

%and includes prefactors which depend on the Coulomb corrected ionization time $t_s^{(1)}$, angle $\phi_v^c$ and momentum $\p^c$. It takes the following form:
%\begin{equation}
%	R_{\kappa l m}(\p) = (-1)^m C_{\kappa l} N_{lm} \sqrt{\kappa} \ \frac{1}{\sqrt{S''(t_s^{(1)}(\p))}} \left[ \ P_l^m\left(\frac{p_z^c}{\p^c+\A(t_s^{(1)})}\right) \ e^{im\phi_v^c(t_s^{(1)}(\p))} \right]
%\end{equation}

So far, we have defined ionization amplitudes asymptotically, projecting onto the free electron states at large times $T$ after the laser pulse has been switched off. However, it is also possible to define subcycle ionization amplitudes. We do this by propagating our electron wavefunction back to some earlier time $t>t_i$ and projecting this onto the plane wave basis $|\p + \A(t)\rangle$. Essentially, this is equivalent to assuming that the long-range potential is negligible from $t\rightarrow T$ and thus ionization is completed by time $t$. In fact, it is possible to show that such subcycle amplitudes take exactly the same form as above. The derivation is found in Appendix C of \cite{circARM1}. That is,
\begin{equation}
	a_{\mathbf{p}}(t) = \langle \p + \A(t) | \Psi(t) \rangle = R_{\kappa l m}(\p) \ e^{-i W_C(t,\p)} \ e^{-i S_{SFA}(t,\p)}. \label{eq:subcycleamp}
\end{equation}
Note that for large times $t=T$ after the laser field has been switched off, $\A(T)=0$ and Eq.\eqref{eq:subcycleamp} reduces to a projection onto field-free states with momentum $\p$ as before. These subcycle ionization amplitudes will allow us to probe the process of ionization in a time-dependent way.

\section{Evaluating the Coulomb correction}

\begin{figure}[h!]
  \begin{center}
    \includegraphics[width=0.35\textwidth]{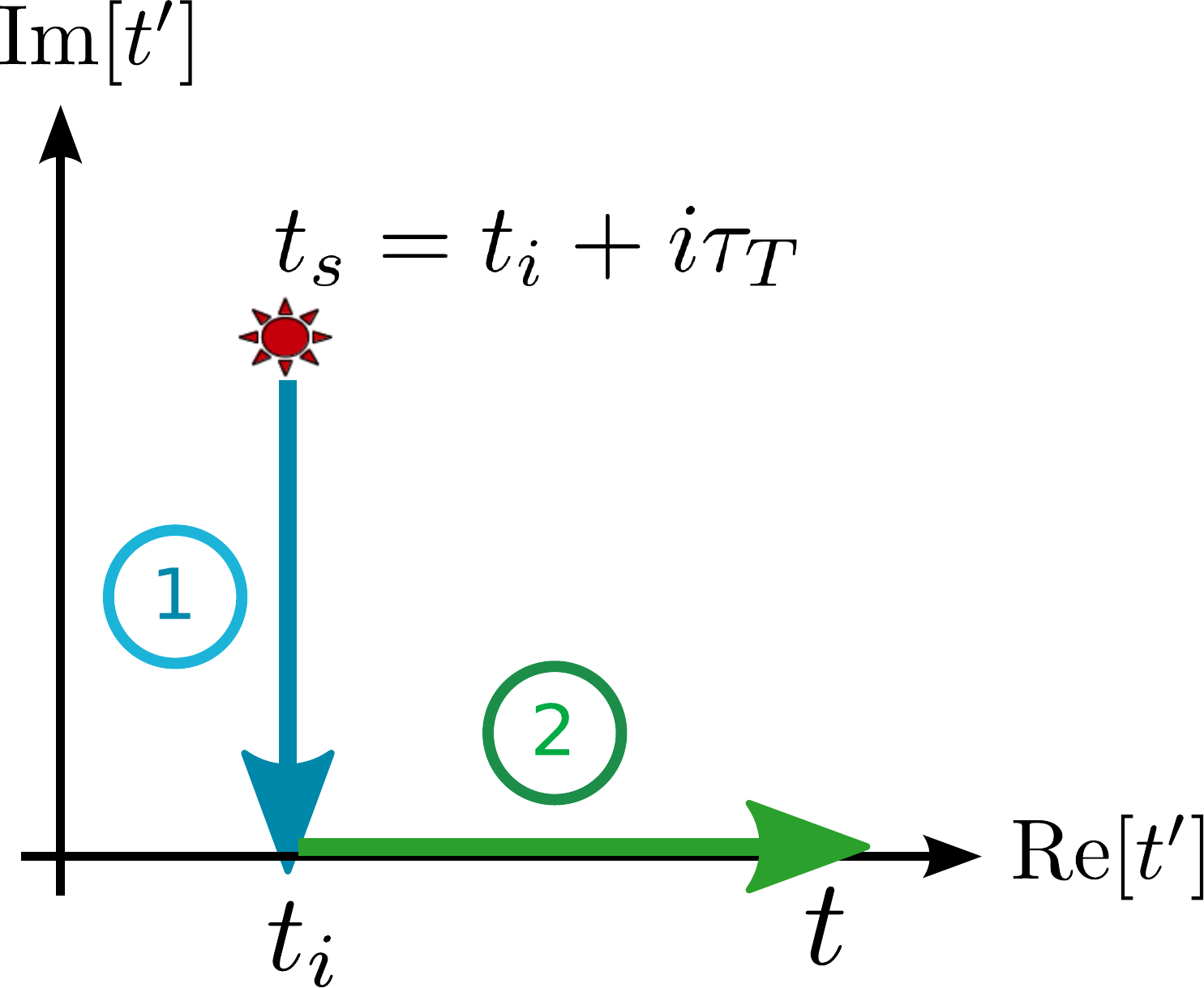}
      \end{center}
  \caption{The integration contour we choose when evaluatating the Coulomb correction term $W_C$ (Eq.\eqref{eq:WC}).}
  \label{fig:contour}
\end{figure}

Let us now turn our attention to the Coulomb correction term $W_C$. We would like to evaluate the integral in Eq.\eqref{eq:WC}. As we did in \cite{ARM1}, we choose our integration contour as follows. First, we integrate parallel to the imaginary axis, down from $\phi_i + i\phi_\kappa$ to $\phi_i$. Next, we integrate along the real axis, from $\phi_i$ to our end point at $\phi_t$ (see Fig.\ref{fig:contour}). That is, we seek to evaluate
\begin{equation}\label{eq:WCsplit}
	W_C = \frac{i}{\w} \int_{\phi_\kappa}^0 d\phi \ U(\r_s(\phi_i + i \phi)) + \frac{1}{\omega} \int_{\phi_i}^{\phi_t} d\phi \ U(\r_s(\phi)).
\end{equation}

\subsection{Analysis of the trajectories}\label{sec:traj}

In the spirit of PPT theory \cite{PPT2}, we can interpret the first leg of this contour as corresponding to the electron's motion under the barrier, while the second leg describes its motion once it emerges in the continuum. Let us consider the trajectories $\r_s(\phi)$ for each of these in turn.

\subsubsection{Under the barrier}

Evaluating $\r_s(\phi_i + i \phi)$ yields the following expressions for the real and imaginary parts of the trajectory for the first leg of the contour:
\begin{align}
	&\r_1(\phi) \equiv \mathrm{Re}[\r_{s,1}] = \frac{F}{\w^2} (\cosh\phi_\tau - \cosh\phi) (\sin\phi_i \hat{\mathbf{x}} - \cos\phi_i \hat{\mathbf{y}}) \\
	& \boldsymbol{\rho}_1(\phi) \equiv \mathrm{Im}[\r_{s,1}] = \left(\frac{F}{w^2} (\sinh\phi_\tau - \sinh\phi) - \frac{p}{\w}(\phi_\tau - \phi)\right) (\cos\phi_i \ \hat{\mathbf{x}} + \sin\phi_i \ \hat{\mathbf{y}})- \frac{p_z}{\w}(\phi_\tau - \phi) \ \hat{\mathbf{z}}.
\end{align}
Note that $\phi$ runs between $\phi_\kappa$ and 0 here. We think of $\phi=\phi_\kappa$ as corresponding to the `entrance' of the barrier and the start of tunnelling, while $\phi=0$ is associated with the barrier `exit', when the electron emerges in the continuum.

Without loss of generality, let us set $p_y$=0. That is, we place our detector along the $x$-axis. In this case, $\phi_i=0$ and the electric field points in the positive $y$-direction at the instant of ionization. If we also set $p_z=0$ (noting that ionization perpendicular to the plane of the laser field is exponentially suppressed), we have
\begin{align}
	&\r_1(\phi,\phi_i=0) = - \frac{F}{\w^2} (\cosh\phi_\tau - \cosh\phi) \ \hat{\mathbf{y}} \\
	& \boldsymbol{\rho}_1(\phi,\phi_i=0) = \left(\frac{F}{w^2} (\sinh\phi_\tau - \sinh\phi) - \frac{p}{\w}(\phi_\tau - \phi)\right) \hat{\mathbf{x}}.
\end{align}
As expected, the real part of the trajectory implies that the electron starts near the origin and escapes in the negative $y$-direction. The imaginary part of the trajectory, on the other hand, varies from positive to negative $x$ values, depending on the drift momentum of the electron $p=p_x$ (that is, the final momentum registered at the detector). Note, however, that the real and imaginary parts are always perpendicular to each other. Note, also, that for the optimal SFA momentum $p=k_0$, the imaginary part of the trajectory vanishes at the barrier exit.

\subsubsection{In the continuum}\label{sec:continuumtraj}

Along the second leg of the integration contour, from $\phi_i$ to $\phi_t$, we have
\begin{align}
	\r_2(\phi) \equiv \mathrm{Re}[\r_{s,2}] &= \left( \frac{F}{w^2} (\sin\phi_i \cosh\phi_\tau - \sin\phi) + \frac{p}{\w} \cos\phi_i(\phi-\phi_i) \right) \hat{\x} \nonumber \\
	+ &\left( -\frac{F}{w^2} (\cos\phi_i \cosh\phi_\tau - \cos\phi) + \frac{p}{\w} \sin\phi_i(\phi-\phi_i) \right) \hat{\y} + \frac{p_z}{\w}(\phi-\phi_i) \ \hat{\mathbf{z}} \\
	\boldsymbol\rho_2(\phi) \equiv \mathrm{Im}[\r_{s,2}] &= \left(\frac{F}{w^2} \sinh\phi_\tau - \frac{p}{\w}\phi_\tau \right) (\cos\phi_i \hat{\mathbf{x}} + \sin\phi_i \hat{\mathbf{y}}) - \frac{p_z}{\w}(\phi_\tau - \phi) \hat{\mathbf{z}}.&
\end{align}

For $p_y=p_z=0$, this simplifies to
\begin{align}
	&\r_2(\phi, \phi_i=0) = \left( -\frac{F}{w^2} \sin\phi + \frac{p}{\w} \phi \right) \hat{\x}
	+ \left( \frac{F}{w^2} (\cos\phi - \cosh\phi_\tau) \right) \hat{\y} \label{eq:r2re} \\
	&\boldsymbol\rho_2(\phi, \phi_i=0) = \left(\frac{F}{w^2} \sinh\phi_\tau - \frac{p}{\w}\phi_\tau \right) \hat{\mathbf{x}}.
\end{align}
From the real part of the trajectory, we see that the electron starts its motion in the continuum at $\r_\mathrm{exit} = \r_s(\phi_i)=-F/\w(\cosh\phi_\tau-1) \hat{\y}$, with initial velocity $\mathbf{v}_i=(p-A_0)\hat{\mathbf{x}}$. We think of this as the exit of the barrier. It then undergoes circular motion superimposed on an overall drift momentum $p$ in the positive $x$-direction. Note that at the moment of ionization, the barrier also rotates in the positive $x$-direction.

The imaginary part of the continuum trajectory, on the other hand, remains constant and equal to its value at the barrier exit. However, with the exception of the optimal momentum $p=k_0$, it is non-zero in general. As we shall see, this non-zero imaginary component will be directly responsible for a shift towards lower momenta in the electron spectra, associated with a deceleration of the electron due to the core potential. It will also lead to small subcycle variations in the total ionization rate in the continuum, which we interpret in terms of the dynamics of electron re-capture and release.

\subsection{Analytical continuation of the Coulomb potential}

As we have seen, the trajectory $\r_s(\phi)$ along both legs of our integration contour is complex in general. Therefore, in order to evaluate $W_C$ (Eq.\eqref{eq:WCsplit}), we must analytically continue our potential $U(\r)$.

Consider the case of the Coulomb potential. For a real vector $\r=(x,y,z)$, we have $V_C(\r)= -Q / \sqrt{\r\cdot\r} = -Q / \sqrt{x^2+y^2+z^2}$. For a complex vector, $\r + i \boldsymbol\rho = (x,y,z) +i(\chi, \eta, \xi)$,
we define
\begin{multline}
	V_C(\mathbf{\r + i \boldsymbol\rho}) = \frac{-Q}{\sqrt{(\r + i \boldsymbol\rho) \cdot (\r + i \boldsymbol\rho)}} = \frac{-Q}{\sqrt{r^2-\rho^2 + 2 i \r\cdot\boldsymbol\rho }} \\
	= \frac{-Q}{\sqrt{(x+i\chi)^2 +(y+i\eta)^2 + (z+i\xi)^2}},
\end{multline}
where $r=|\r|^2$ and $\rho=|\boldsymbol\rho|^2$. It is easy to verify that this satisfies the Riemann-Cauchy conditions. We shall choose the branch cut of the complex square root to be infinitesimally above the negative real axis. In doing so, we ensure that our contour never crosses a branch cut.

If we let
\begin{equation}
	a = r^2-\rho^2 \quad \text{and} \quad b = 2\r\cdot\boldsymbol\rho,
\end{equation}
we can explicitly express the Coulomb potential in terms of its real and imaginary parts as follows:
\begin{equation}
	V_C = -\frac{Q}{\sqrt{2(a^2+b^2)}} \left( \sqrt{\sqrt{a^2+b^2} + a}
	- i \ \mathrm{sgn}(b) \sqrt{\sqrt{a^2+b^2} - a} \right),
\end{equation}
where we define
\begin{equation}
	\mathrm{sgn}(b) =
		\begin{cases}
			-1, \quad \text{for } b \leq 0 \\
			+1, \quad \text{for } b > 0.
		\end{cases}
\end{equation}

\begin{figure}[h!]
  \begin{center}
    \includegraphics[width=0.82\textwidth]{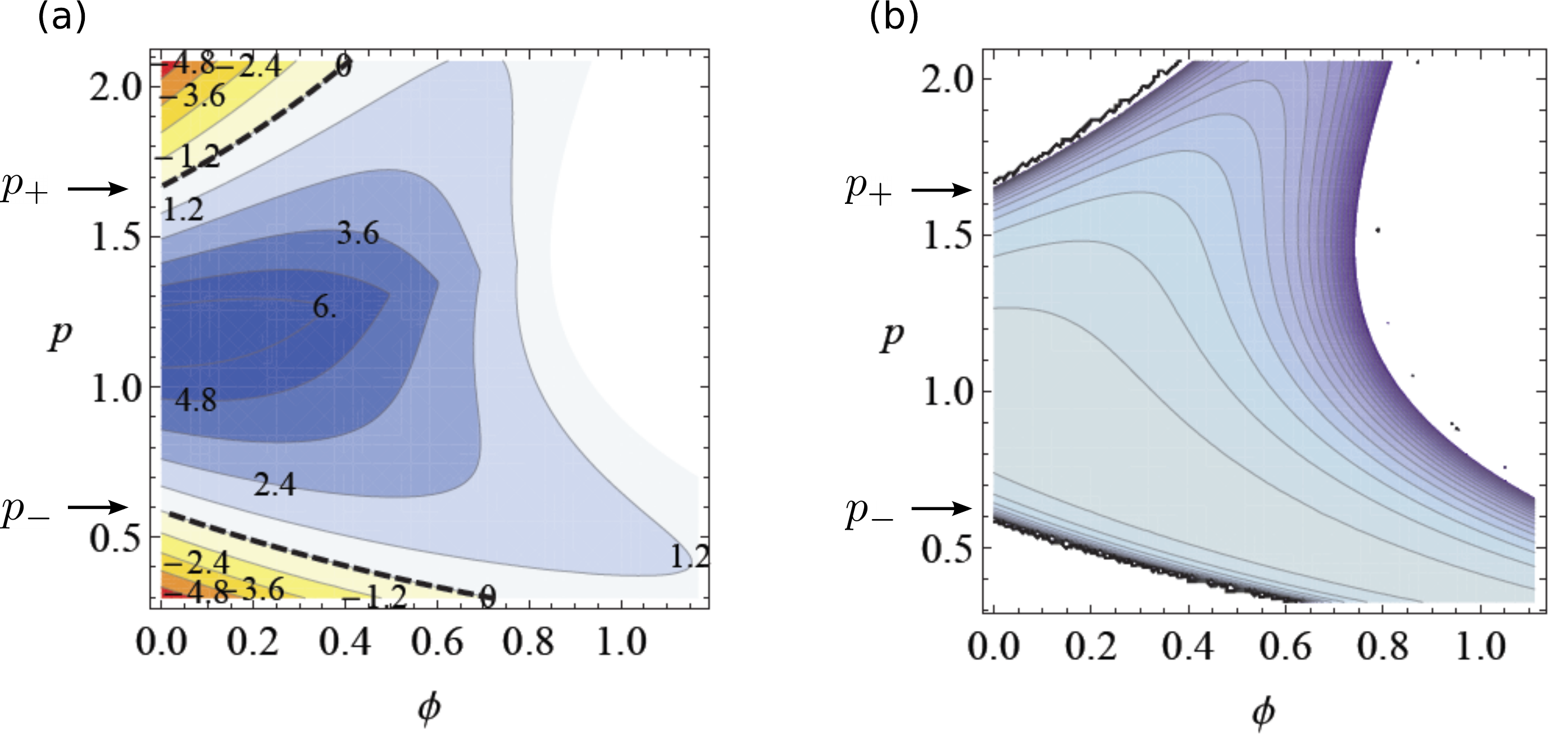}
      \end{center}
  \caption{(a) The difference between the magnitude of the real and imaginary part of the trajectory under the barrier $r_1-\rho_1$, as a function of imaginary time $\phi$ (in radians) and drift momentum $p$ (in a.u.) in the $x$-direction. In the blue region, $r_1>\rho_1$ and $V_C$ is purely real. In the red/yellow region, $r_1<\rho_1$ and $V_C$ is purely imaginary. The black dashed lines correspond to $r_1=\rho_1$, where $V_C$ is singular. Note that $\phi$ evolves from $\phi_\kappa$ to zero (right to left) as the electron moves away from the atom.
    (b) The real part of the Coulomb potential under the barrier. It is singular along the lines $r_1=\rho_1$ and zero where $r_1<\rho_1$.}
  \label{fig:underthebarrier}
\end{figure}

Note that in addition to the singularity at the origin that we are familiar with ($r=\rho=0$), we now have an additional singularity when $\r\cdot\boldsymbol\rho = 0$ and $r=\rho$. This is an important case to consider since, as we have seen, $\r\cdot\boldsymbol\rho = 0$ everywhere under the barrier. Figure \ref{fig:underthebarrier} shows the difference between the magnitudes of the real and imaginary parts of the under-barrier trajectory $r_1-\rho_1$, as a function of imaginary time $\phi$ and momentum $p$. For momenta near the centre of the SFA momentum distribution $p \in (p_-, p_+)$, we see that $r_1>\rho_1$ everywhere. $V_C$ therefore remains real and there is no singularity along our integration contour. However, for momenta outside of this central region, we will pass over the line $r_1=\rho_1$ as time evolves from $\phi_\kappa$ to the ionization time $\phi_i=0$. In this case, we encounter a singularity and $V_C$ changes from real to imaginary as the electron tunnels. Fortunately, unlike the singularity at the origin, this singularity is integrable and the integrals for $W_C$ always converge. Note that at $p^-$ and $p^+$, defined by $r(p_\pm,\phi_i)=\rho(p_\pm,\phi_i)$, the singularity occurs at the barrier exit.

\subsection{Evaluating the Coulomb phase}\label{sec:evaluatingWC}

Since we are interested in studying the effect of the Coulomb correction term on electron spectra and ionization rates, the quantity of interest is the ionization probability $|a_\p(t)|^2 \sim \exp[2 \mathrm{Im}S_{SFA}] \exp[2 \mathrm{Im}W_C]$. We shall therefore focus on the imaginary part of the Coulomb phase. That is, we are interested in $\mathrm{Im}[W_C] = W_{C_1} + W_{C_2}$, where
\begin{align}
	&W_{C_1} = \frac{1}{\w} \int_{\phi_\kappa}^0 d\phi \ \mathrm{Re}[V_C(\r_{s,1})] \\
	&W_{C_2} = \frac{1}{\w} \int_{\phi_i}^{\phi_t} d\phi \ \mathrm{Im}[V_C(\r_{s,2})].
\end{align}

\begin{figure}[h!]
  \begin{center}
    \includegraphics[width=0.5\textwidth]{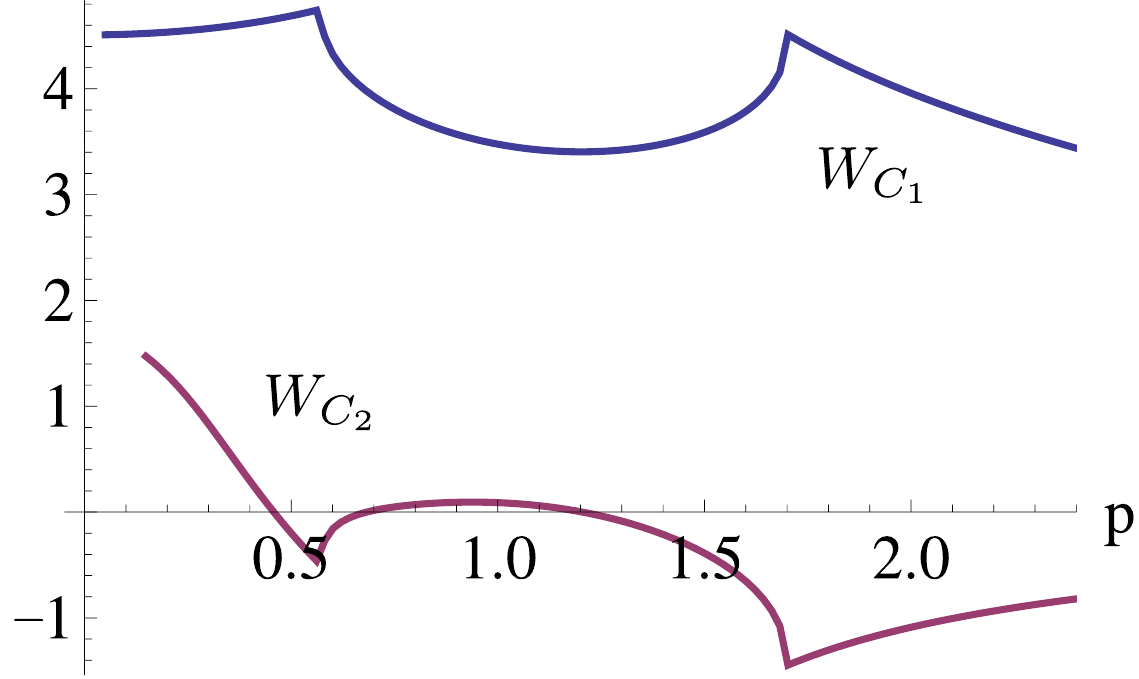}
      \end{center}
  \caption{The Coulomb correction terms $W_{C_1}$ and $W_{C_2}$ plotted against the momentum registered at the detector $p=\sqrt{p_x^2+p_y^2}$, for $\lambda=800$nm,
  $I=2.6 \times 10^{14}$W/cm$^2$, $I_p=0.5$.
  $W_{C_2}$ is shown at time $\phi_t=20$.}
  \label{fig:WC1WC2}
\end{figure}

Figure \ref{fig:WC1WC2} shows these Coulomb correction terms as a function of the drift momentum of the electron, that is, the final momentum measured at the detector. Overall, we see that the under-barrier Coulomb term $W_{C_1}$ is always positive and hence leads to an enhancement in ionization rate. The continuum Coulomb correction $W_{C_2}$ can lead to either enhancement or suppression, depending on the momentum. The two sharp points we see in each of the two graphs correspond to $p^-$ and $p^+$, the momenta at which the integrable $r=\rho$ singularity occurs at the exit of the barrier. However, note that these sharp points precisely cancel each other out when we compute the total Coulomb correction $\mathrm{Im}[W_C] = W_{C_1}+W_{C_2}$, which is the actual quantity of interest.

\begin{figure}[h!]
  \begin{center}
    \includegraphics[width=1\textwidth]{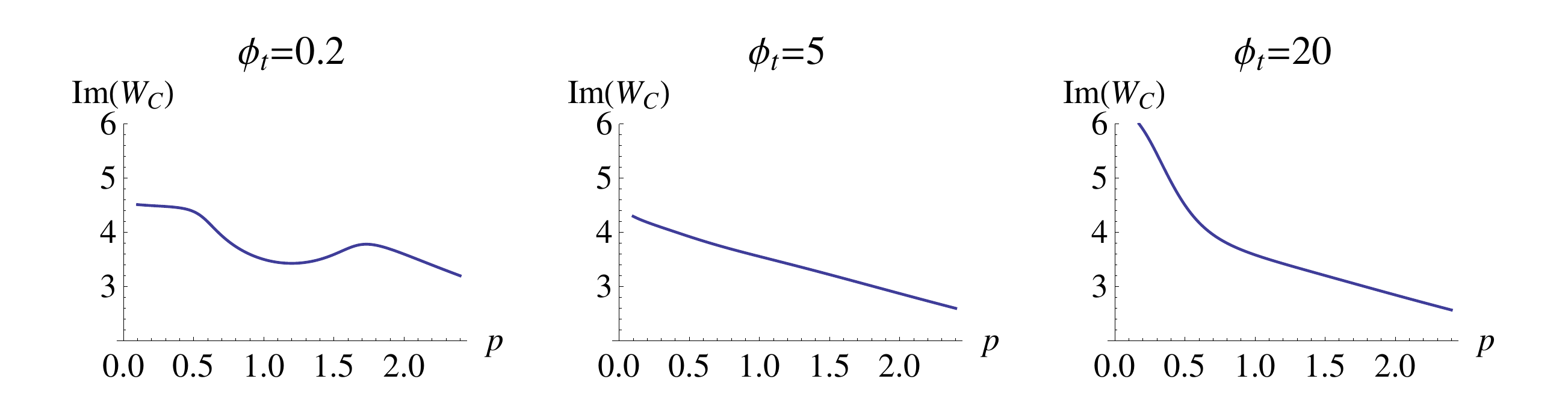}
      \end{center}
  \caption{The total Coulomb correction $\mathrm{Im}[W_C]=W_{C_1}+W_{C_2}$, plotted against momentum for three different observation times, for $\lambda=800$nm, $I=2.6 \times 10^{14}$W/cm$^2$, $I_p=0.5$.}
  \label{fig:WC}
\end{figure}

Figure \ref{fig:WC} shows the total Coulomb correction $\mathrm{Im}[W_C]$, plotted against the drift momentum of the departing electron for a series of three different observation times. Notice that it is smooth everywhere and varies with time. Notice, also, that at sufficiently large times, $\mathrm{Im}[W_C]$ increases with decreasing momenta. As we shall see, this will give rise to a shift in the electron momentum distribution. We investigate the effect of this term in greater detail in the next section.

%Notice that our analysis has naturally yielded a Coulomb correction term that is both time and momentum dependent. It is this fact that will allow us to study the way in which the core potential modifies electron spectra, and how ionization rates build up over time. As we shall see, both these effects come about as a consequence of the imaginary parts of the continuum trajectories in our theory.

\section{Coulomb Effects: Electron Spectra and Subcycle Ionization Rates}

Armed with our time and momentum dependent Coulomb correction, we are now ready to evaluate the ionization probability $|a_\p(t)|^2 \sim \exp[2 \mathrm{Im}S_{SFA}] \exp[2 \mathrm{Im}W_C]$. We shall compare this to the standard SFA result $\exp[2\mathrm{Im}S_{SFA}]$, where the effect of the core is neglected. Doing this, we find two main effects:
\begin{enumerate}
	\item The peak in the electron distribution shifts towards lower momenta. This shift is accumulated as the electron travels in the continuum.
	\item In addition to the well-known increase in the total ionization rate acquired under the barrier, we observe further small variations after the electron emerges in the continuum.
	%	The total ionization rate increases. While this increase is primarily acquired under the barrier, we find further small variations after the electron emerges in the continuum.
\end{enumerate}

\subsection{Electron spectra: A shift to lower momenta}

Since our Coulomb correction is momentum dependent, our theory allows us to study the effect of the core on electron spectra. For a given observation angle, we investigate the shape of $|a_p(t)|^2 \sim \exp[2 \mathrm{Im}S_{SFA}(p)] \exp[2 \mathrm{Im}W_C(p,t)]$ as a function of canonical momentum $p$. Note that canonical momentum here always refers to the momentum measured asymptotically at the detector, even though we evaluate $|a_p(t)|^2$ subcycle. As mentioned before, we can think of $p$ as the electron's drift momentum as it moves in the presence of the laser field.

Recall, first, the SFA case, $|a^{SFA}_p|^2 \sim \exp[2 \mathrm{Im}S_{SFA}]$. As discussed in section \ref{sec:amplitudes}, the momentum distribution in this instance is a Gaussian-like function centred at $p=k_0$, where $k_0$ is given by Eqn.\eqref{eq:k0}. Short wavelengths, weak field strengths and small ionization potentials correspond to smaller $k_0$.

%Let us first briefly review the SFA case (without the Coulomb correction): $|a^{SFA}_p(t)|^2 = e^{2 \mathrm{Im}[S_{SFA}]}$. In this instance, we know that the momentum distribution is a Gaussian-like function centred at $p=k_0$, where $k_0$ is given by Eqn.\eqref{eq:k0}. In the adiabatic limit $\gamma \ll 1$, we have $k_0 \rightarrow A_0$. Recalling that velocity at the time of ionization is $|\mathbf{v}(t_i)|=|\p+\A(t_i)|=|p-A_0|$, we see that this limit corresponds to the electron appearing in the continuum with zero initial velocity. For larger $\gamma$, however, the peak moves to progressively larger values $k_0>A_0$ and the electron emerges with an initial velocity in the direction in which the barrier is rotating. This is a non-adiabatic effect \cite{Barth2011}.

\begin{figure}[h!]
  \begin{center}
    \includegraphics[width=0.75\textwidth]{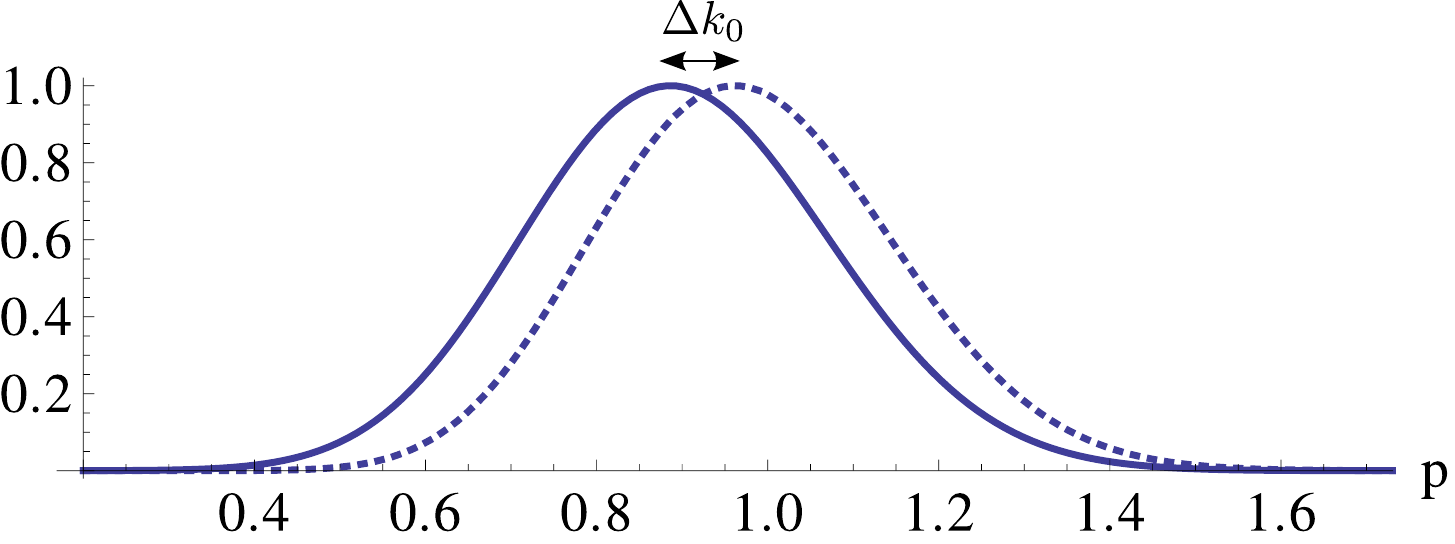}
      \end{center}
  \caption{Electron momentum distribution with (solid) and without (dotted) the Coulomb correction, normalised to a height of 1, for $\lambda=600$nm, $I=2.6 \times 10^{14}$W/cm$^2$ and $I_p=0.5$, evaluated at an asymptotic time $\phi_t=1000$.}
  \label{fig:spectra600}
\end{figure}

\begin{figure}[h!]
  \begin{center}
    \includegraphics[width=1\textwidth]{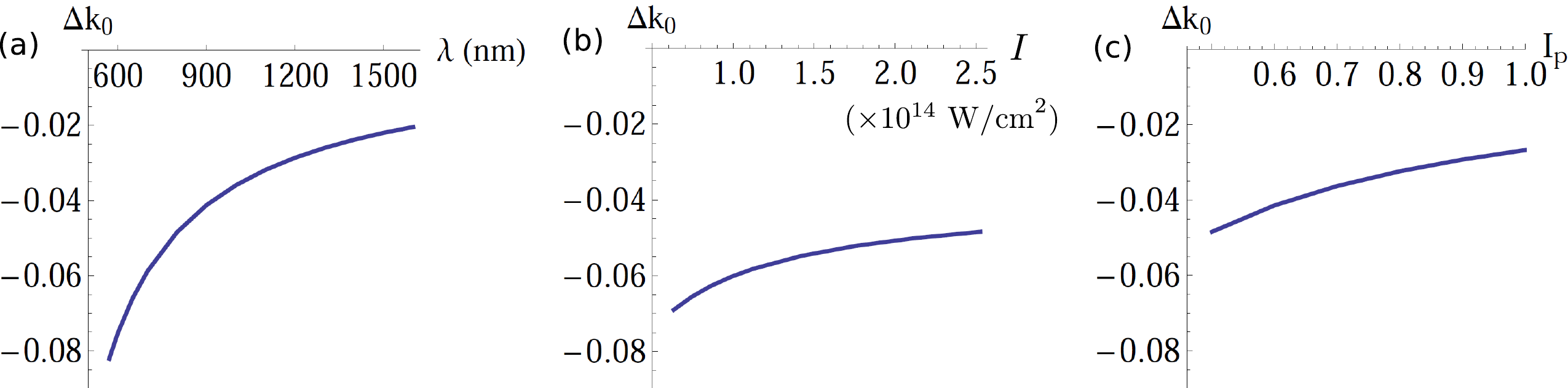}
      \end{center}
  \caption{The shift of the peak of the photoelectron spectrum due to the Coulomb correction as a function of (a) wavelength, (b) field intensity, and (c) ionization potential, evaluated at an asymptotic time $\phi_t=1000$.}
  \label{fig:shifts}
\end{figure}

As alluded to in section \ref{sec:evaluatingWC}, if we now include the Coulomb correction factor, we find that the distribution shifts towards low momenta. Figure \ref{fig:spectra600} shows an example of this shift in the photoelectron spectrum, evaluated at large asymptotic time $\phi_t=1000$. Figure \ref{fig:shifts} shows how this shift varies with the system parameters. Notice that we observe the largest shifts at short wavelengths, weak fields and low ionization potentials, that is, when the SFA distribution is centred at low momenta.

\begin{figure}[h!]
  \begin{center}
    \includegraphics[width=0.5\textwidth]{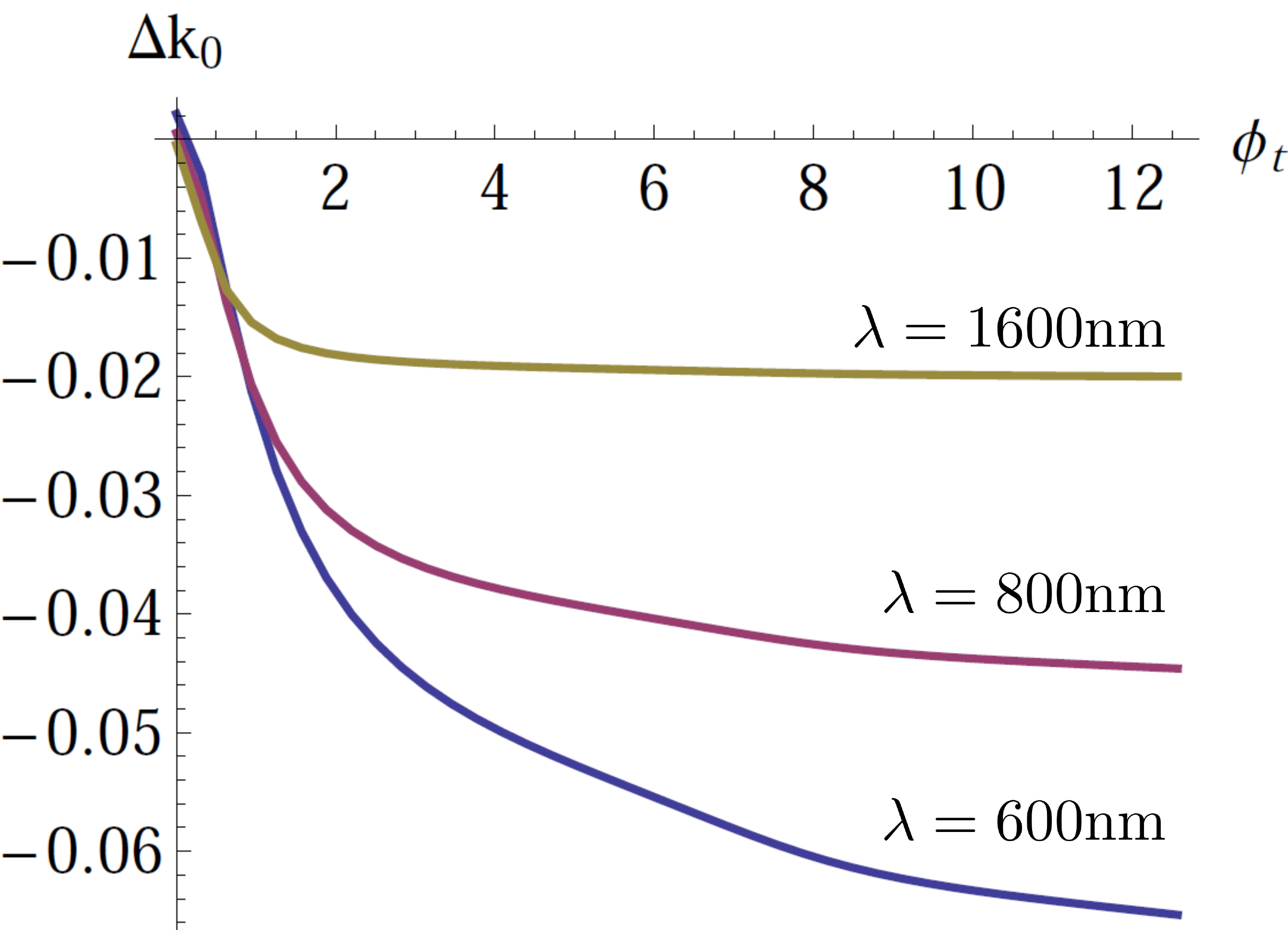}
      \end{center}
  \caption{The shift of the peak of the photoelectron spectrum as a function of time in the continuum $\phi_t$.}
  \label{fig:shiftsvstime}
\end{figure}

Since our theory is time-dependent, we can go further and investigate the dynamics associated with these shifts in a time-resolved way. Doing this, we find that the shift in $|a_p(t)|^2$ is predominantly accumulated while the electron travels in the continuum. Immediately after ionization, at the barrier exit, the peak is very close to the SFA prediction. However, as we allow time to evolve from $\phi_i$ along the real time axis, the peak moves towards lower momenta and soon approaches its final asymptotic value. Figure \ref{fig:shiftsvstime} shows this subcycle variation for three different wavelengths. Note that for shorter wavelengths (associated with a smaller initial drift momentum $k_0$), the peak takes a longer time to reach its final asymptotic value.

Clearly, it is natural to interpret these shifts as the deceleration of the electron wavepacket by the core as it moves away from its parent atom or molecule. As discussed in the introduction, this is a well-recognised effect, which is commonly treated by propagating classical trajectories within a two step model. Within our theory, however, we find that it emerges completely naturally, without adding anything ad hoc, and comes about as a direct consequence of the imaginary components of the continuum trajectories. Had the continuum trajectories been purely real, we would not observe any such shifts. The imaginary parts of the continuum trajectories are therefore not merely a mathematical curiosity, but encode an important and real physical effect. 

We can quantify the shift of the peak in the electron spectrum by differentiating $|a_p(t)|^2$ with respect to $p$. Solving $\frac{\partial}{\partial p} [\mathrm{Im}[S_{SFA}(k_C)] + \mathrm{Im}[W_C(k_C)]] = 0$, we find the Coulomb-corrected peak momentum $k_C = k_0 + \Delta k_0$. We can estimate this by expanding about $k_0$ to first order in the Coulomb correction as follows,
\begin{align} 
	\Delta k_0^{(1)}(t) &= \frac{-\mathrm{Im}\left[\left.\frac{\partial}{\partial p} W_C\right|_{k0} \right]}{\mathrm{Im}\left[\left.\frac{\partial^2}{\partial p^2}S_{SFA}\right|_{k_0}\right]}\\
	&=
	\frac{1}{\phi_\tau(k_0)} \left( \int_{\phi_\kappa}^0 d\phi \left.\frac{\partial}{\partial p} U(\r_{s,1}(\phi,p))\right|_{k_0} - U(\r_{s,1}(\phi_\kappa,k_0)) \left.\frac{\partial \phi_\tau}{\partial p}\right|_{k_0} \right. \nonumber \\ 
	 &\left. \qquad \qquad \qquad + \int_{\phi_i}^{\phi_t} d\phi \ \mathrm{Im}\left[\left.\frac{\partial}{\partial p} U(\r_{s,2}(\phi,p))\right|_{k_0} \right] \right). \label{eq:deltak0approx}
\end{align}
Note that the first two terms, which come from the under-barrier contribution $W_{C_1}$, approximately cancel each other, and as a result, the shift is primarily accumulated after the electron emerges in the continuum (the final term in Eq.\eqref{eq:deltak0approx}).

\subsection{Total ionization rate: non-adiabatic effects and subcycle variations}

It is well-established that accounting for the Coulomb potential leads to an overall enhancement in the total ionization rate predicted by SFA. This can be understood by considering the shape of the tunnelling barrier. Compared to the short-range potential implicit within the strong field approximation, the barrier for a Coulomb potential is smoother and lower, making it easier for the electron to tunnel through. In the adiabatic limit, this increase is typically approximated using \cite{PPT2,Keldysh65,ADK}
\begin{equation}\label{eq:adiabaticrate}
	\frac{w}{w^{SFA}} \approx \left( \frac{2 \kappa^3}{F} \right)^{2/\kappa}.
\end{equation}
Within our ARM formalism, which is valid well into the non-adiabatic regime and enables us to study ionization dynamics in a time-resolved way, we are able to go beyond this result.

To do this, we calculate the total ionization rate in a given direction by integrating over momenta $p$,
\begin{equation} \label{eq:rateintegral}
	w(t) = \int dp \ |a_p(t)|^2 \propto \int dp \ e^{2 \mathrm{Im}[S_{SFA}]} e^{2 \mathrm{Im}[W_C]}.
\end{equation}
We shall again fix $p_z=0$ here, noting that ionization in the $z$-direction is exponentially suppressed. Without loss of generality, we also set $p_y=0$ and integrate over $p=p_x$. Since our field is circular and we consider long pulses, the problem is symmetric and this calculation yields the ionzation rate for any given direction in the plane of laser polarization.

\subsubsection{Ionization rates via the saddlepoint method: a first approximation}

In our preceeding paper \cite{circARM1}, we approximated the above integral using the saddlepoint method, relying on the fact that the SFA term is a relatively narrow Gaussian-like function centred at $k_0$. Within this approximation, to lowest order in $W_C$, we obtain
\begin{equation}\label{eq:saddlepointrate}
	w \approx \frac{\sqrt{\pi}}{\sqrt{|\mathrm{Im}[\frac{\partial^2}{\partial p^2} S_{SFA}(k_0)]|}} e^{2 \mathrm{Im}[S_{SFA} (k_0)]} e^{2 \mathrm{Im}[W_C (k_0)]}.
\end{equation}
Here, the correction to the SFA ionization rate is simply $\exp[2 \mathrm{Im} W_C(k_0)]$, the Coulomb term evaluated along the optimal trajectory where $p=k_0$. This result coincides with the non-adiabatic Coulomb correction in PPT \cite{PPT3}. Note that in contrast to the purely static Coulomb correction in Eq.(\ref{eq:adiabaticrate}), $W_{C_1}$ now also depends on wavelength.

Recalling from section \ref{sec:traj} that the optimal trajectory is purely real at real times, we see that $W_{C_2}(k_0)$ is zero. Hence, within this approximation, the Coulomb modification of the total ionization rate comes only from the under-barrier part $W_{C_1}$, and no further changes to the total rate occur as we evolve in real time. This result seems quite intuitive. The total ionization rate depends only on what happens to the electron while it is tunnelling, and we can say that ionization is completed at the ionization time $t_i$ when the electron appears in the continuum. As plausible as this picture seems, however, it does not tell the whole story.

\subsubsection{A more accurate calculation: frequency dependence and time-resolved ionization rates}

In order to obtain a more accurate description of the total subcycle ionization rates, we evaluate the integral in Eq.\eqref{eq:rateintegral} numerically and divide this by the SFA rate $w^{SFA} = \int dp \ e^{2\mathrm{Im}[S_{SFA}]}$. In doing so, we find deviations from the result obtained using the saddlepoint method. In particular, our Coulomb correction to the ionization rate acquires a time-dependence, and no longer remains constant in the continuum.  

\begin{figure}[h!]
  \begin{center}
    \includegraphics[width=0.5\textwidth]{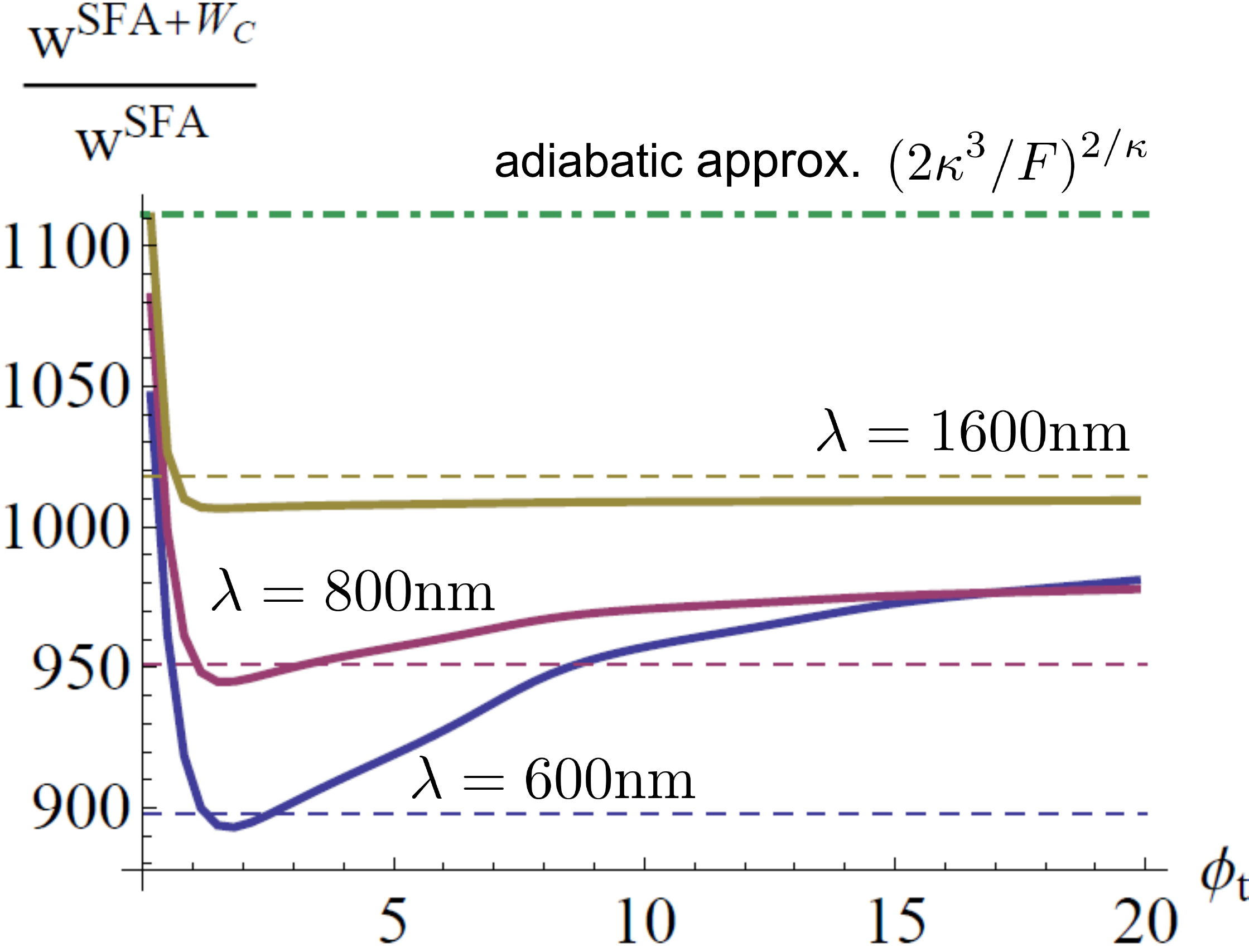}
      \end{center}
  \caption{Coulomb enhancement of the total ionization rate as a function of time, normalized by the SFA rate. Solid lines are the results of numerical integration. Dashed lines show the lowest order saddlepoint approximation (that is, the non-adiabatic Coulomb correction $\exp[W_{C_1}(k_0)]$). The dot-dashed green line indicates the commonly used adiabatic approximation given by Eq.\eqref{eq:adiabaticrate}.}
  \label{fig:medtimerates}
\end{figure}

Figure \ref{fig:medtimerates} shows these time-resolved results for $\lambda=$ 600nm, 800nm and 1600nm. In all cases, we find an initial decrease in the rate on a timescale of one quarter of a laser cycle or less. We suggest that this could be attributed to the Coulomb trapping of the departing electron into Rydberg states. At shorter wavelengths, as we enter the nonadiabatic $\gamma \gtrsim 1$ regime, we find that this initial decrease is followed by a gradual increase in rate over a longer period of time, where the extra contribution comes from small momenta. This may be explained by electrons being shaken back out of Rydberg states at higher frequencies. Again, note that these subcycle variations arise as a direct consequence of the imaginary parts of the continuum trajectories. That is, the imaginary components again give rise to non-trivial physical effects.

\begin{figure}[h!]
  \begin{center}
    \includegraphics[width=0.5\textwidth]{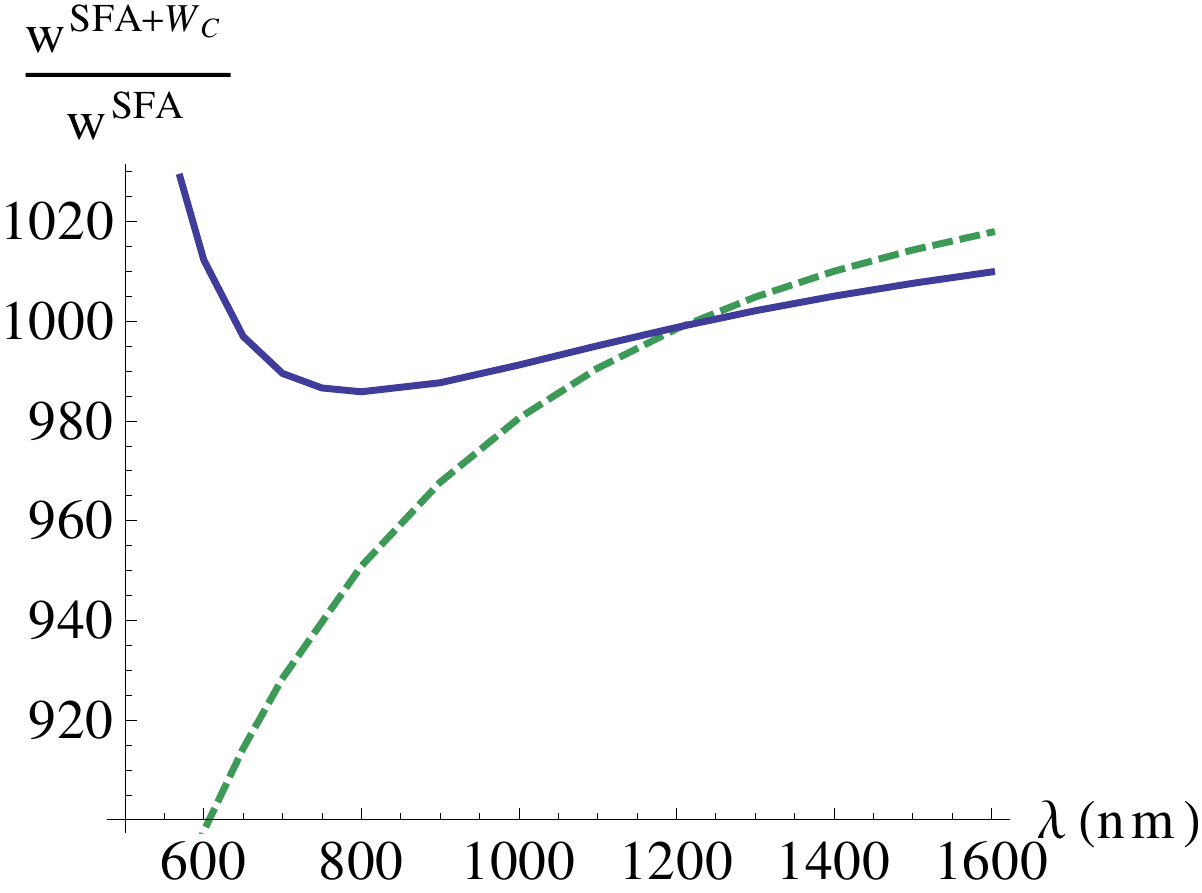}
      \end{center}
  \caption{The Coulomb enhancement of the total ionization rate, calculated using the saddlepoint approximation (green dashed) and by evaluating the integral in Eq.\eqref{eq:rateintegral} exactly (blue solid), as a function of wavelength.}
  \label{fig:ratevswavelength}
\end{figure}

The innaccuracy of the lowest order saddlepoint approximation for the integral over $p$ in Eq.\eqref{eq:rateintegral} comes about due to the shift of the optimal momentum $\Delta k_0$ and the re-shaping of the photoelectron distribution. The deviation between the two as a function of wavelength is shown in Figure \ref{fig:ratevswavelength}.

\section{Conclusion}

In this work, we have studied the effect of the core potential in strong field ionization by circularly polarized fields, using ionization amplitudes derived via the time-dependent analytical R-matrix approach \cite{ARM1,ARM2,circARM1}. Our theory applies well into the non-adiabatic regime, and the Coulomb correction is both momentum and time dependent, which has allowed us to investigate its effect on electron spectra and subcycle ionization rates in a time-resolved way.

Within this approach, we have been naturally led to the concept of complex trajectories. With the exception of the optimal trajectory associated with momentum $k_0$, our trajectories have a non-zero imaginary part all the way to the detector. We have shown that these imaginary components, in fact, give rise to real physical effects.

In electron spectra, we find that the Coulomb term shifts the SFA distribution towards low momenta. This effect is accumulated after the ionization time, as the electron travels in the continuum and time evolves along the real axis. It is natural to interpret this as the deceleration of the electron wavepacket by the core as it moves away from its parent atom or molecule. As we would expect, the observed shift is largest when the electron emerges with a smaller drift velocity and hence spends more time in the vicinity of the core.

We also find that our correction leads to subcycle variations of the ionization rate in the continuum. This is in addition to the well-known enhancement of the SFA rate accumulated while the electron is tunnelling. We interpret these variations in terms of electron recapture into Rydbery states and their subsequent re-release due to interaction with the laser field.

Both of these effects come about as a direct consequence of the imaginary parts of the continuum electron trajectories in our theory.

\section{Acknowledgments}
We thank M. Ivanov for useful discussions, and D. Bauer for useful comments. L.T. and O.S. gratefully acknowledge the support of the Leibniz graduate school DiNL. J.K. and O.S. gratefully acknowledge the support of the Marie Curie ITN CORINF.

\end{document}